\documentclass[pra, 10pt, notitlepage, twocolumn]{revtex4-1}
\usepackage{graphicx}
\usepackage[caption=false]{subfig}
\usepackage{amssymb}
\usepackage{amsmath}
\usepackage{amsfonts}
\usepackage{floatrow}
\usepackage{mathtools}
\usepackage[export]{adjustbox}
\usepackage{tikz}
\usetikzlibrary{matrix,arrows,decorations.pathmorphing}

\def\dbar{{\mathchar'26\mkern-12mu d}}

\newcommand{\braket}[2]{\left\langle #1 | #2 \right\rangle}
\newcommand{\proj}[1]{| #1\rangle\!\langle #1 |}

\newcommand{\Tr}{\mathrm{Tr}}
\newcommand{\Det}{\mathrm{Det}}

\newcommand{\inv}{\text{-1}}
\newcommand{\T}{\mathsf{T}}
\newcommand{\rank}{\mathrm{rank}}
\newcommand{\R}{\mathbb{R}}

\newcommand{\pref}[1]{(\ref{#1})}

\begin{document}

\title{Non-holonomic tomography I:  The Born rule as a connection between experiments}
\author{Christopher Jackson and Steven van Enk}
\affiliation{
Oregon Center for Optical Molecular and Quantum Sciences\\
Department of Physics\\
University of Oregon, Eugene, OR 97403}

\begin{abstract}
In the context of quantum tomography, we recently introduced a quantity called a partial determinant \cite{jackson2015detecting}.
PDs (partial determinants) are explicit functions of the collected data which are sensitive to the presence of state-preparation-and-measurment (SPAM) correlated errors.
As such, PDs bypass the need to estimate state-preparation or measurement parameters individually.
In the present work, we suggest a theoretical perspective for the PD.
We show that the PD is a holonomy and that the notions of state, measurement, and tomography can be generalized to non-holonomic constraints.
To illustrate and clarify these abstract concepts, direct analogies are made to parallel transport, thermodynamics, and gauge field theory.
This paper is the first of a two part series where the second paper\cite{nonholo2} is about scalable generalizations of the PD in multiqudit systems,
with possible applications for debugging a quantum computer.
\end{abstract}

\maketitle

\section{Introduction}

In quantum computing, a recent problem has been learning how to estimate quantum gates
while taking into account that there are small but significant errors in the states prepared and measurements made to probe such gates,
so called SPAM errors \cite{merkel}.
Several works have come out to solve this, \cite{merkel,gst,stark}, all of which speak to the notion of a ``self-consistent tomography.''
These works also make an important common assumption: that the uncontrolled fluctuations in the SPAM are not correlated.
So in \cite{jackson2015detecting} the obvious question was asked: what if the states and measurements made were actually correlated with each other?

Even though this question can be asked for classical systems, this is an especially interesting question for quantum systems.
Standard quantum theory tells us that reality articulates itself as discrete events.
The probabilities of these events are further understood to be the product of \emph{two} things: a state and a set of possible outcomes.
More precisely, the Born rule in its modern form tells us that the distribution of these events is the inner product of a density operator and a POVM.
This is what makes a quantum theory distinct from a classical one as it allows for fundamental randomness because a state is no longer an outcome in itself:
state and outcome become distinct notions.
As distinct as these notions are, they are nevertheless inseparable because each quantum event measured is always and only the product of a state and a possible outcome
| a fact which is especially apparent in tomography.
Put another way, there is no quantum state defined operationally, independent of the resource of known possible outcomes and vice versa.

This brings up a fundamentally important point which is that the concept of states and observables as separate and independent is a subjective or man-made distinction,
reflecting the model of standard quantum theory.
In the presence of SPAM correlations, average state and average measurement parameters cannot be defined as statistically independent quantities,
consistent with all possible state and measurement settings.
However, one can still define average state and average measurement parameters \emph{locally} over the space of device settings.
A simple but subtle example of such locally defined quantities can be found in thermodynamics
| the caloric, ``$Q$'', and potential energy, ``$W$'', represented by inexact heat and work forms which sum to changes in the energy, $dU = \dbar Q + \dbar W$,
which is globally defined over the thermodynamic state space.
A more standard example can be found in quantum electrodynamics
| the electron kinetic momentum, $-i D_\mu$, and the photon vector potential, $A_\mu$, which sum to the canonical momentum, $-i\partial_\mu = -i D_\mu + A_\mu$,
globally defined over position space.

In order to illustrate these analogies explicitly, we will consider a toy analogy to quantum tomography with SPAM errors.
This toy model replaces the state and measurement with single parameters, which can be correlated.
We demonstrate precisely how the toy analog of the partial determinant from \cite{jackson2015detecting}
has the same structure as $\oint\dbar Q$ from thermodynamics or $\oint A\cdot dx$ from QED.
Such ``loop'' integrals are generally called holonomies and the forms they integrate can be referred to as non-holonomic constraints.
Finally, we translate these results to actual quantum tomography, completing the perspective of non-holonomic quantum tomography.

\section{State-Preparation, Measurement, their Correlation, and Data}

\subsection{The Born Rule and Tomography}

One could say that the Born rule was originally, since the 1920s, used exclusively to \emph{predict} the distributions of events \emph{from} states and observables.
Standard textbook treatments will denote the Born rule by $P(s|\psi) = |\braket{s}{\psi}|^2$, thus introducing the notions of state and measurement outcome.
Statistical observables\footnote{
There is a slight conflict of language here as modern field theorists like to use ``observable'' to refer to cross-sections, lifetimes, etc. which we refer to as ``statistical observables'' as opposed to ``quantum observables'' which refer to operators in a theory and what we mean throughout this paper by ``observable.''}
are then calculated from classical probability theory and typical expressions like
\begin{widetext}

\begin{equation}\label{typical}
\langle s \rangle = \sum_s s P(s) = \sum_{s,\psi} s P(s|\psi)P(\psi) = \Tr\left(\sum_{\psi} P(\psi)\proj{\psi}\right)\left(\sum_{s} s \proj{s}\right)= \Tr\rho\Sigma = \langle\Sigma\rangle
\end{equation}

\end{widetext}
appear, introducing the notions of a classically mixed state and a quantum observable.
Since distinct quantum systems can interact, the notion of an ancilla can be introduced and measurements can be generalized from an orthonormal basis to a positive operator valued measure (POVM).

In more recent years, the Born rule has found a different application in so called quantum state tomography \cite{raymer, paris}, where states are concluded from the distribution of measured events and various known POVMs.
After this, it was quickly recognized that the Born rule could just as well be used for so called detector tomography \cite{lundeen, feito}, where POVM elements are concluded from the distribution of events and known states.
It had even been demonstrated that one could perform state tomography through unknown POVMs from other known states with a technique similar to applying the Born rule twice, bypassing the need to parameterize unknown POVMs \cite{cooper2014local}.

Any application of the Born rule where \emph{both} state preparation and measurement are unknown \cite{merkel,gst,stark} we will henceforth refer to as SPAM tomography.
The central feature which makes SPAM tomography distinct from other tomographies is the presence of gauge degrees of freedom.
In this case, state and measurement parameters are explicitly inseparable because the Born rule cannot uniquely determine them individually from the statistics alone.
Work has been done to recover unique estimates for individual state and measurement parameters \cite{stark} under further assumptions.
Of course, such work also makes the implicit assumption that there are no correlated SPAM errors.

\begin{widetext}

\begin{figure}[h!]
\centering
\includegraphics[height=1.75in]{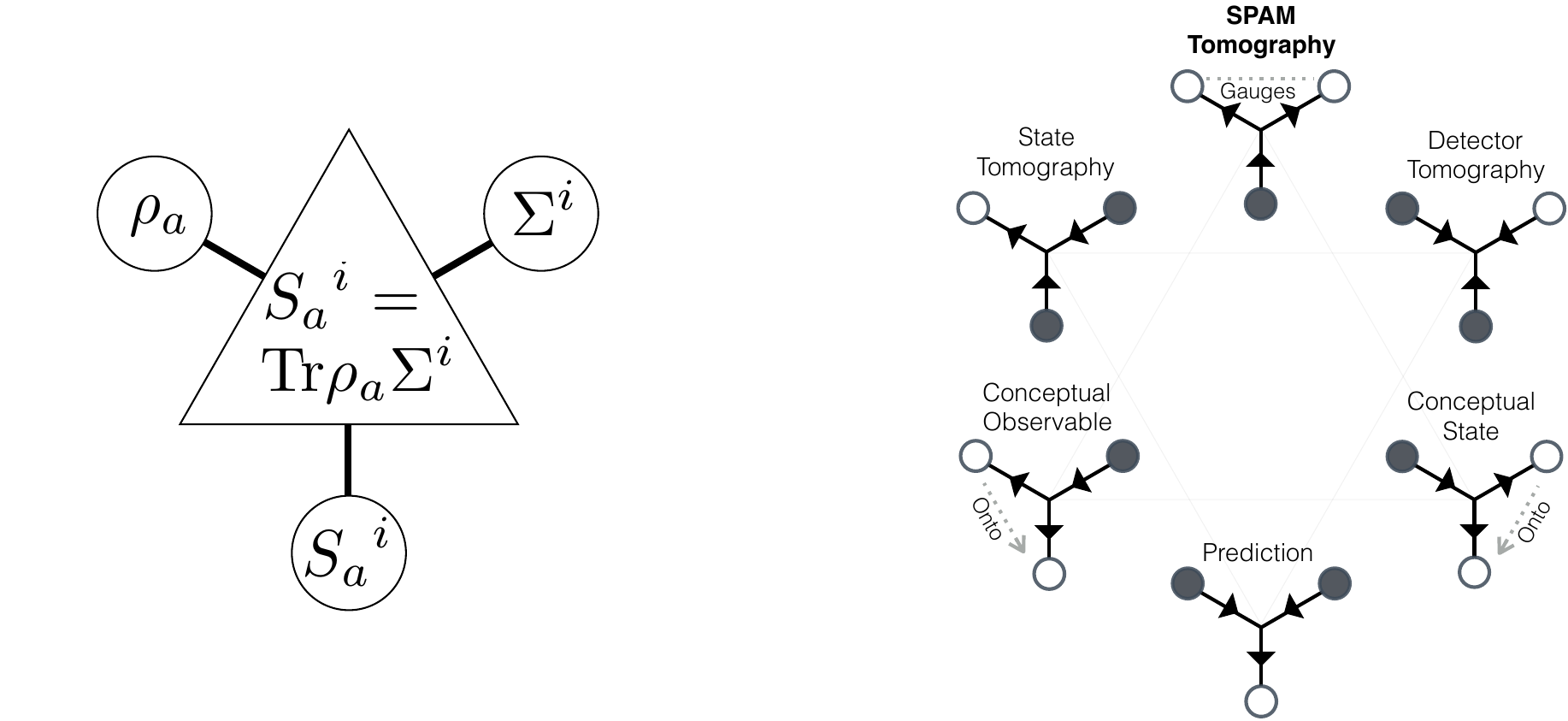}
\caption{
Schematic diagram for the various perspectives of the Born rule.
Left: In the most general sense the Born rule is simply a constraint between states ($\rho_a$), observables ($\Sigma^i$), and data (${S_a}^i$).
Right: The six perspectives of the Born rule | states, observables, and data are represented spatially as in the left diagram and darker corners represent parameters that are fixed externally.
``Prediction'' fixes states and observables to conclude measured data.  ``State Tomography'' uses fixed observables and data to conclude states.  ``Detector Tomography'' uses fixed states and data to conclude observables.
Dual to ``State Tomography'' is the ``Conceptual State'', where a fixed state is understood as a map from observables onto their `expectation value'.
Dual to ``Detector Tomography'' is the ``Conceptual Observable'', where a fixed observable is understood as a map from states onto their `expectation value'.
Finally, dual to ``Prediction'' is ``SPAM Tomography'', where state-observable relationships are concluded from fixed data.
}\label{triad}
\end{figure}

\end{widetext}

In the context of our work, where we \emph{do} allow for correlated SPAM errors, a crucial point must be made concerning our use of the $\langle\rangle$ notation.
On the leftmost side of Equation (\ref{typical}), $\langle\rangle$ refers to the expectation value of a random variable, $s$.
On the rightmost side of Equation (\ref{typical}), $\langle\rangle$ loses this meaning as it does \emph{not} refer to the expectation value of an operator, $\Sigma$, but rather an inner product of $\Sigma$ with the density operator.
In both cases, the \emph{distribution of quantum events is completely attributed to the state} and this assumption is perhaps further obscured by Dirac's bra-ket notation.
For our purposes in SPAM tomography, we will not use $\langle\rangle$ in this way, beyond Equation (1).
Rather, $\langle\rangle$ will refer to an expectation value where states and observables are themselves considered random variables.
Specifically, if $\rho$ is a density operator representing the state and $E$ is a POVM element representing a possible outcome, then one must understand that
\begin{equation}
	f = \langle\mathrm{Tr}\rho E\rangle
\end{equation}
where $f$ is an estimate of the probability (obtained from the frequency of the measured outcome) and $\langle\rangle$ is the average over the ensemble of trials.
The measured frequency is SPAM correlated if
\begin{equation}\label{wtf}
	\langle\mathrm{Tr}\rho E\rangle \neq \mathrm{Tr} \langle \rho \rangle \langle E \rangle.
\end{equation}
We will examine such correlations in a much simpler context in the next section but details and examples may also be found in \cite{jackson2015detecting}.

\subsection{A Toy Example}\label{toy}

The problem of whether states and measurements are correlated is fundamentally interesting
because states and observables are not individually accessible in principle by experiment alone.
In other words, the quantities of the right hand side of Equation \pref{wtf} cannot be measured without arbitrarily well characterized devices.
Nevertheless, it was demonstrated in \cite{jackson2015detecting} that there is still a way to detect such correlations using properties of the data alone,
bypassing the need to estimate state and observable parameters separately.
The basic essence of that result can be illustrated by the following toy problem:

\begin{figure}[h!]
\centering
\includegraphics[height=0.75in]{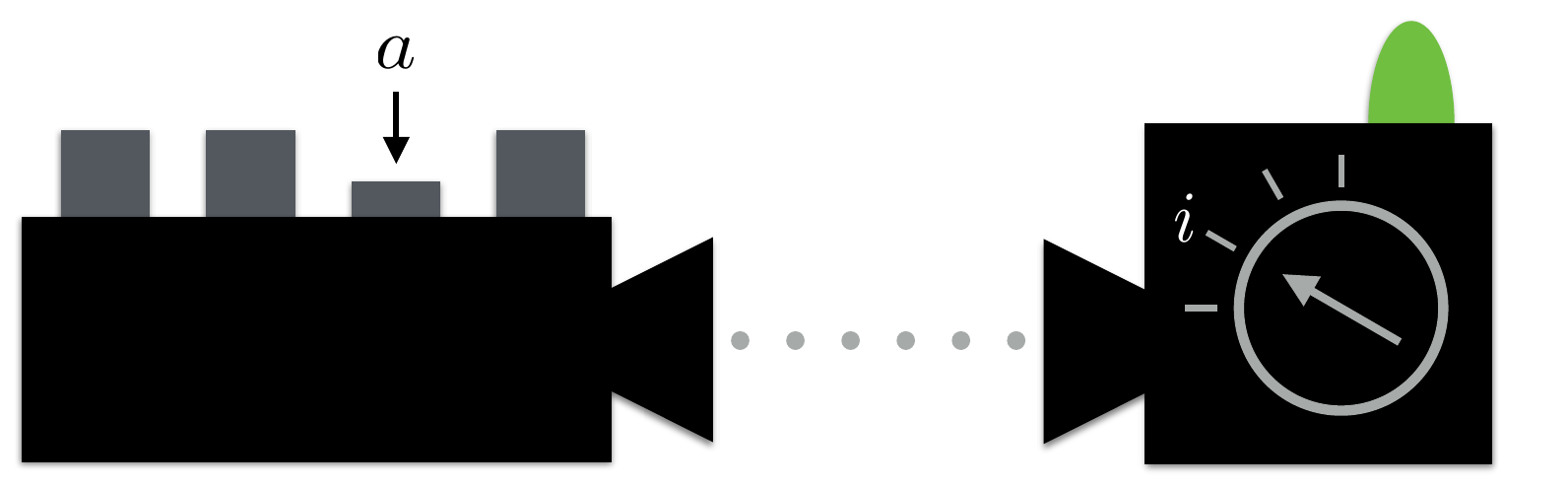}
\caption{
On the left is a device which prepares various signals on demand depending on which button, $a\in\{1,\ldots,N\}$, is pressed.
On the right is a device which blinks to indicate a signal with a certain property depending on which setting, $i\in\{1,\ldots,M\}$, a dial is turned to.
}\label{knobs}
\end{figure}

Consider a device with various settings, $a$, each of which prepare a different signal on demand by the press of a button.
Consider also a detector with various settings, $i$, each of which detect a particular property of the signal indicated by the blink of a light.
Now suppose it is suspected that each setting of the preparation device actually produces the same signal,
only that each setting produces the signal with varying probabilities of success, $p_a$.
Suppose further that the detection device is expected to simply indicate the presence of the signal, only that each setting can register a signal with varying probabilities of success, $w^i$.
We then imagine that $p_a$ and $w^i$ are actually unknown and that we are only able to change settings and record whether the light blinks or not.

Let us indicate by ${f_a}^i$ the measured frequency with which the light blinks when the devices are set to $(a,i)$.
If one can assume that the performance of the devices and their settings are uncorrelated, then one can simply identify (after many runs of the experiment)
\begin{equation}
	{f_a}^i = p_a w^i.
\end{equation}
However, relaxing this assumption to allow for the possibility of correlations,
one must be more careful about the quantities defined so to make the more general identification that
\begin{equation}
	{f_a}^i = {\langle pw \rangle_a}^i
\end{equation}
where we have introduced notation ${\langle \rangle_a}^i$ to represent the average over the ensemble of trials for the pair of settings $(a,i)$.
The subtlety here is that the devices can still be represented by single parameters, $p$ and $w$,
only now these parameters are to be understood as random variables which fluctuate depending on the setting $(a,i)$.

The presence of SPAM correlation is simply when the frequencies, ${\langle pw \rangle_a}^i$ (which are what we have access to) are such that
\begin{equation}
	{\langle pw \rangle_a}^i \neq \langle p \rangle_a\langle w \rangle^i.
\end{equation}
It would seem that to identify such a circumstance one would have to measure $\langle p \rangle_a$ and $\langle w \rangle^i$ individually.
However, such measurements would require devices which are already well characterized, unlike the devices we have.
What we would like to do is detect if such correlations are present between our devices given only our humble, imperfect, uncalibrated devices.

\subsection{Gauge Degrees of Freedom}

In such a situation, one must acknowledge that there will always be so called gauge degrees of freedom.
If one was given the promise that a pair of device parameters were in fact SPAM uncorrelated,
then there would still be a one-parameter family of possible values for the average state parameter and average detector parameter.
Specifically, for a possible pair of values $\big(\langle p\rangle,\langle w\rangle\big)$ such that $\langle pw \rangle = \langle p\rangle\langle w\rangle$,
the pair $\big(g\langle p\rangle, g^\inv\langle w\rangle\big)$ is just as possible.\footnote{
Of course, $g$ must has a compact range so that the interpretation of \unexpanded{$\big(g\langle p\rangle, g^\inv\langle w\rangle\big)$} as a pair of probabilities still makes sense.
However, this detail is not of concern for this paper.}
If the devices are SPAM uncorrelated over a range of settings $a\in\{1,\ldots,N\}$ and $i\in\{1,\ldots,M\}$,
then the set of possible average values continue to define exactly one gauge parameter.
This is perhaps best illustrated by observing ${\langle pw \rangle_a}^i = \langle p \rangle_a\langle w \rangle^i$ as a matrix equation,
\begin{equation}\label{globaleff}
\left[
\begin{array}{ccc}
	{\langle pw \rangle_1}^1	&&	{\langle pw \rangle_1}^M	\\
		&	\ddots	&	\\
	{\langle pw \rangle_N}^1	&&	{\langle pw \rangle_N}^M	
\end{array}
\right]
=
\left[
\begin{array}{c}
	\langle p \rangle_1	\\
	\vdots	\\
	\langle p \rangle_N	
\end{array}
\right]
\left[
\begin{array}{ccc}
	\langle w \rangle^1	&	\cdots &	\langle w \rangle^M
\end{array}
\right],
\end{equation}
so that if $\left(\left[\langle p \rangle_1 \cdots \langle p \rangle_N\right]^\T,\left[\langle w \rangle^1 \cdots \langle w \rangle^M\right]\right)$ is possible,
then so is $\left(g\left[\langle p \rangle_1 \cdots \langle p \rangle_N\right]^\T,g^\inv\left[\langle w \rangle^1 \cdots \langle w \rangle^M\right]\right)$.

To handle this gauge degree of freedom, it is useful to define the following notion:
The collected data, ${\langle pw \rangle_a}^i$, for a pair of devices is \emph{effectively} (SPAM) uncorrelated if Equation \pref{globaleff} exists
| that is, if the experimentally accessible left-hand side can be expressed as in the right-hand side for some $\left[\langle p \rangle_a\right]^\T$ and $\left[\langle w \rangle^i\right]$.
Considered as a matrix, $D = [{\langle pw \rangle_a}^i]$, one should recognize that this definition is equivalent to an upper bound on the rank, $\rank(D)\le1$.
Such a bound on the rank can be further quantified by considering the determinant of every $2\times2$ submatrix of the data, so called ($2\times2$) minors.
Specifically, every such minor must be zero if the data is effectively uncorrelated.
One should recognize that such conditions are properties of the data collected by just our humble devices alone.

Having mentioned some standard notions from linear algebra, there is an alternative set of notions which support the same analysis.
These notions are also more geometric in their perspective, which one might have suspected to exist from the association of gauge.
The technique which accompanies these notions further has an obvious tomographic interpretation.
As a final statement of this prelude, the alternative technique we are referring to is also what generalizes to actual quantum tomography.

\subsection{Partial Determinants}\label{PD}

To demonstrate, we will need only to consider two settings per device, $N=M=2$.
For simplicity, let us denote the quantities ${\langle pw \rangle_1}^1$, ${\langle pw \rangle_1}^2$, ${\langle pw \rangle_2}^1$, and ${\langle pw \rangle_2}^2$
by simply $\langle pw \rangle$, $\langle pv \rangle$, $\langle qw \rangle$, and $\langle qv \rangle$, respectively and refer to them as \emph{data}.
Further, let us denote the settings $(1,1)$, $(1,2)$, $(2,1)$, and $(2,2)$ respectively as $(p,w)$, $(p,v)$, $(q,w)$, and $(q,v)$ and refer to them as \emph{experiments}. 

If one considers only the measured quantity ${\langle pw \rangle}$, then such a datum is always effectively uncorrelated and should thus be associated with a gauge degree of freedom.
This is true as well for the other data, $\langle pv \rangle$, $\langle qw \rangle$, and $\langle qv \rangle$ considered individually.
Considering these data individually means that they have the property of being (effectively) uncorrelated \emph{locally}, in which case each experiment should be understood to correspond to a local gauge degree of freedom.
Explicitly, ``local'' is relative to the space of experimental settings which here consists only of 4 points
(though we will consider the continuous case soon enough in the next section.)

Each gauge degree of freedom is arbitrary in the sense that they cannot be defined without the resource of better calibrated devices.\footnote{
One can argue that such devices do not exist other than by assumption!}
However, these gauge degrees of freedom are still related to each other because the experiments can share common settings.
For example, let us parameterize the gauge of the experiment $(p,w)$ with ${\langle w \rangle}$ and the gauge of $(p,v)$ with ${\langle v \rangle}$.
Since these two experiments share the setting $p$, their corresponding gauge degrees of freedom are related by the data
because ${\langle v \rangle} = \frac{\langle pv \rangle}{\langle pw \rangle} {\langle w \rangle}$.
In other words, the data can be interpreted as a \emph{connection} between the gauge of each experiment.
The connection is itself not uniquely determined by the data, but this is only because it is intimately related to the gauge
| e.g. if we had instead parameterized the the gauge of experiment $(p,w)$ with ${\langle p \rangle}$,
then the above connection would have been rather ${\langle v \rangle} = {\langle pv \rangle}/{\langle p \rangle}$.

The gauge of each experiment $(p,w)$ represents the fact that the corresponding data $\langle pw \rangle$ is locally (effectively) uncorrelated.
Nevertheless, it may still be the case that the data of all four experiments is not (effectively) uncorrelated \emph{globally} so that one may not be able to write
\begin{equation}\label{global}
D = 
\left[
\begin{array}{cc}
	\langle pw \rangle & \langle pv \rangle	\\
	\langle qw \rangle & \langle qv \rangle
\end{array}
\right]
=
\left[
\begin{array}{c}
	\langle p \rangle	\\
	\langle q \rangle
\end{array}
\right]
\left[
\begin{array}{cc}
	\langle w \rangle & \langle v \rangle
\end{array}
\right]
\end{equation}
simultaneously.
As observed earlier, such data is globally (effectively) uncorrelated if and only if $\det D = 0$.
Assuming $\langle pw \rangle\langle qv \rangle\neq0$, the $\det D = 0$ condition is equivalent to
\begin{equation}
	\Delta(D) \equiv \frac{\langle pv \rangle\langle qw \rangle}{\langle pw \rangle \langle qv \rangle}=1
\end{equation}
and it is this quantity which generalizes to the full quantum problem.\cite{jackson2015detecting}
Since $\Delta$ is only a function of data, it is manifestly gauge invariant.
$\Delta$ is called a \emph{partial determinant} because of the analogy to the above problem and because it is not generally a single number, but rather a matrix of reduced size ($d^2 \times d^2$ for $d$-dimensional Hilbert spaces.)

Restating the (toy) result,
\begin{verse}\centering
	$D$ is globally (effectively) uncorrelated if and only if $\Delta(D)=1$,
\end{verse}
The reader may be familiar with a proof of this using the language of standard linear algebra\footnote{
Remember that this result is equivalent to the commonly known property that $D$ is rank ($\le$) 1 if and only if $\Det D = 0$.}
(considering $D$ as an operator and considering its null space, etc.)
However to emphasize the perspective, we include here a more tomographic proof:
The ``only if'' can be proved by simple substitution.
For the ``if'' direction, one first remembers that they can always choose $\langle p \rangle$ and $\langle w \rangle$ such that $\langle pw \rangle = \langle p \rangle \langle w \rangle$.
Having chosen $\langle p \rangle$ and $\langle w \rangle$, one may then fix $\langle q \rangle = \langle qw \rangle/\langle w \rangle$ and $\langle v \rangle = \langle pv \rangle/\langle p \rangle$.
Notice that this fixing of $\langle q \rangle$ and $\langle v \rangle$ is analogous to state and detector tomography.
Finally, if $\Delta(D) = {\langle pv \rangle\langle qw \rangle}{\langle pw \rangle \langle qv \rangle}=1$
then $\langle qv \rangle = \langle pv \rangle\langle qw \rangle/\langle pw \rangle = \langle q \rangle \langle w \rangle$, which finishes the proof.

\begin{figure}[h!]
\centering
\includegraphics[height=2in]{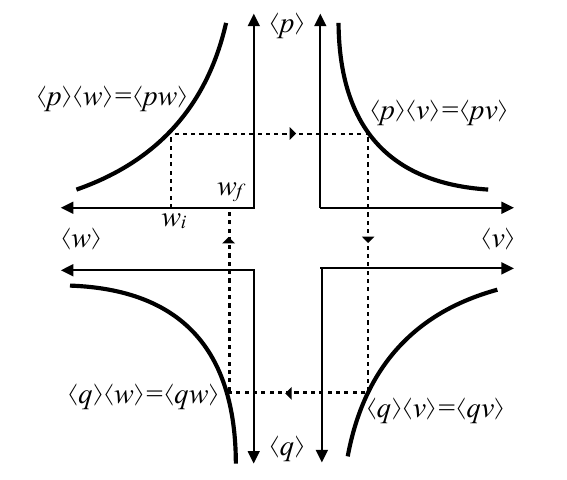}
\caption{Illustration of the PD as a Holonomy:
Each experiment $(p,w)$ has a local gauge degree of freedom because it is effectively SPAM uncorrelated, $\langle p \rangle \langle w \rangle = \langle pw \rangle$.
The data $\langle pw \rangle$ further provides a connection between adjacent gauge degrees of freedom by the assumption that they share independent settings.
Such a connection defines a non-holonomic constraint when $w_f = \frac{\langle pv \rangle\langle qw \rangle}{\langle pw \rangle \langle qv \rangle}w_i \neq w_i$.
A particular $w_i$ fixes the gauge which can either represent an arbitrary choice or some external information.
The PD $\Delta=\frac{w_f}{w_i}$ is gauge invariant.}\label{demo}
\end{figure}

Summarizing, we have developed a perspective for analyzing toy data which parallels a perspective for analyzing quantum data:
Considering the settings $(p,w)$, $(q,w)$, $(p,v)$, and $(q,v)$ as individual experiments,
these settings act as \emph{coordinates} for the space of experiments so that one can say, for example,
experiments $(p,w)$ and $(q,w)$ are displaced from each other by keeping the measurement setting constant.
Further, each individual experiment is \emph{effectively uncorrelated} because we can always choose $\langle p \rangle$ and $\langle w \rangle$ such that $\langle pw \rangle = \langle p \rangle \langle w \rangle$.
The freedom of that choice is a \emph{gauge} degree of freedom and is further a \emph{local} one because each experiment has this property.
Finally, there is a \emph{connection} between the gauges of each experiment
because we can write equations like $\langle p \rangle = \langle pw \rangle/\langle w \rangle$ | 
that is, a choice of $\langle w \rangle$ fixes the gauge of experiment $(p,w)$ which consequently fixes the gauge of experiment $(p,v)$.
With this connection, the partial determinant has the interpretation of performing tomography in a loop,
with a value which measures a contradiction (see Figure \ref{demo}), reflecting the presence of SPAM correlation.
Cast in this language, we have demonstrated that a PD is a \emph{holonomy}.
We shall proceed to explain this further.
At last, it is the tomographic interpretation of this holonomy which is why we refer to any analysis with PDs \emph{non-holonomic tomography}.

\section{Holonomy}\label{horono}

\begin{figure}[h!]
\centering
\includegraphics[height=1.25in]{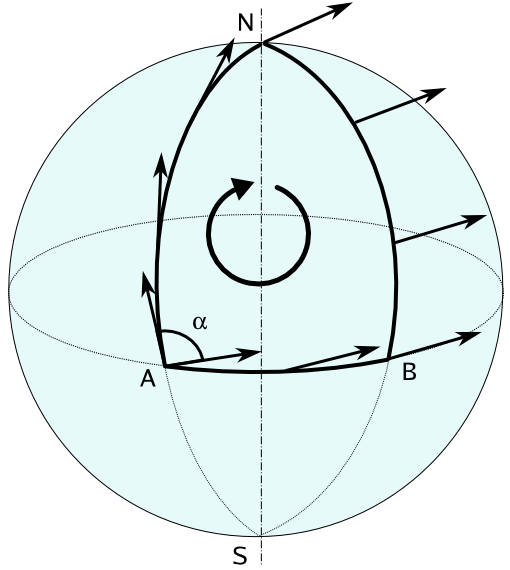}
\caption{Probably the most familiar example of holonomy is the parallel transport of a tangent vector on the sphere.}\label{sphereho}
\end{figure}

Holonomy is a concept which has become quite ubiquitous in modern physics and mathematics.
Applications range from geometric phases to Yang-Mills Lagrangians, all of which share the notion of a non-holonomic constraint.
Perhaps the simplest physical examples of non-holonomic constraint are the thermodynamic concepts of heat and work,
although thermodynamics is typically not considered in this way.
The simplest mathematical example is probably parallel transport through a sphere,
where a tangent vector will turn with an angle proportional to the solid-angle subtended by the loop traversed (Figure \ref{sphereho}.)

Characteristic of these non-holonomic systems are local degrees of freedom (such as heat or angle) whose differential can be integrated over contours defined within certain dimensions (such as the thermodynamic state or the point on a sphere.)
However, these integrals will have non-zero values over \emph{closed} contours,
reflecting that these local degrees of freedom cannot be globally defined as additional dimensions like the ones which defined the contour.
Such integrals are called holonomies and their non-zero values may be interpreted as a measure of contradiction
or inability to integrate the local degree of freedom to a global coordinate.

The technical notion of heat as a holonomy is not standard and so an elaboration is in order.
This will allow us to draw an analogy from which the perspective of non-holonomic tomography will be more explicit.
Using the language of gauges in such a non-standard way, it will also be appropriate to relate these notions to their more familiar application in gauge field theory.
After having established theses connections (no pun intended) we will then rewrite non-holonomic tomography in this field theoretic language.
For completeness, we include a section on the actual quantum analogue of the toy problem to make all the respective technical aspects clear.

\subsection{Analogy: Thermodynamics}\label{OMGthermo}

\begin{widetext}

\begin{figure}[h!]
\centering
\includegraphics[width=5in]{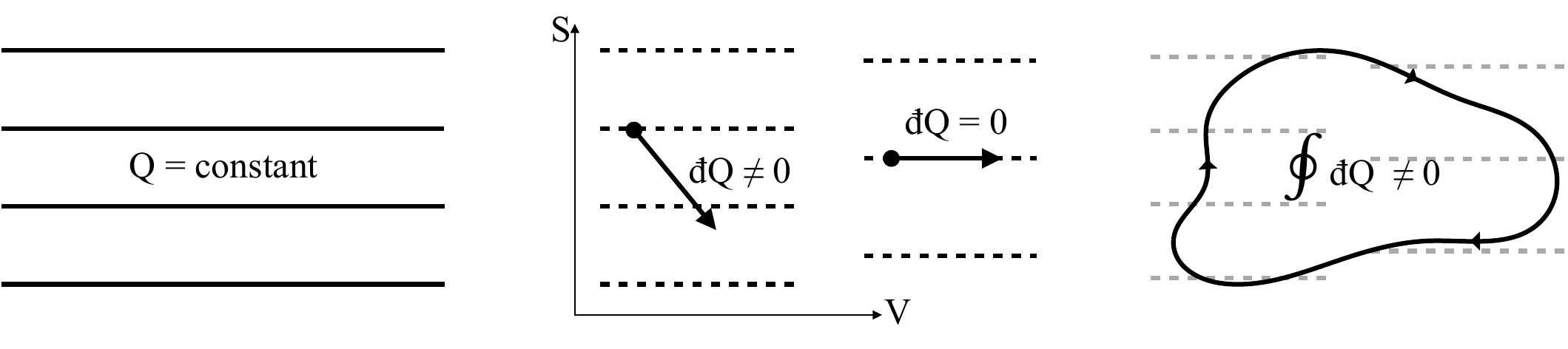}
\caption{Left: Holonomic constraints can be written globally and therefore used as coordinates.
Middle: Non-holonomic constraints are only local and cannot define coordinates.
The dashed lines are supposed to convey that a notion of ``transverse'' is still present
but the distance between the layers of constraint can be correlated with coordinates along the layers.
Right: Non-holonomic constraints thus give rise to holonomies or non-zero integrals over closed contours.}\label{nonholo}
\end{figure}

\end{widetext}

For a thermodynamic system such as an ideal piston, the notion of an adiabatic process can be defined but cannot be extended to a notion of heat as a quantity.
This is because heat can be transferred (into other forms of energy) over closed loops in state space (see Figure \ref{nonholo}.)
This transfer of heat is the holonomy and the integrals $\int_\gamma \dbar Q$ are the connection.
Put another way, the connection $\int_\gamma \dbar Q$ can be thought of as a change in some quantity (like caloric), $\Delta Q$,
but only locally because one can have nonzero changes in the heat upon a \emph{return} to the same state.

However, the notions of energy and entropy do exist as globally defined state variables and heat can be thought of as the energetic response generated by changes in entropy,
\begin{equation}
	\dbar Q = TdS.
\end{equation}
The coefficient of response is the temperature which can depend on other degrees of freedom within the state space, such as volume:
\begin{equation}
	T(S,V) = \left.\frac{\partial U}{\partial S}\right|_V
\end{equation}
This extra dependence on other degrees of freedom is what makes $\,\dbar Q$ non-holonomic, non-integrable, or inexact (words which are synonymous in this context.)
For such a temperature that depends on volume, one could say that the energy transfer generated by a fixed displacement in entropy is \emph{correlated} with the volume.

\begin{figure}[h!]
\centering
\includegraphics[height=0.75in]{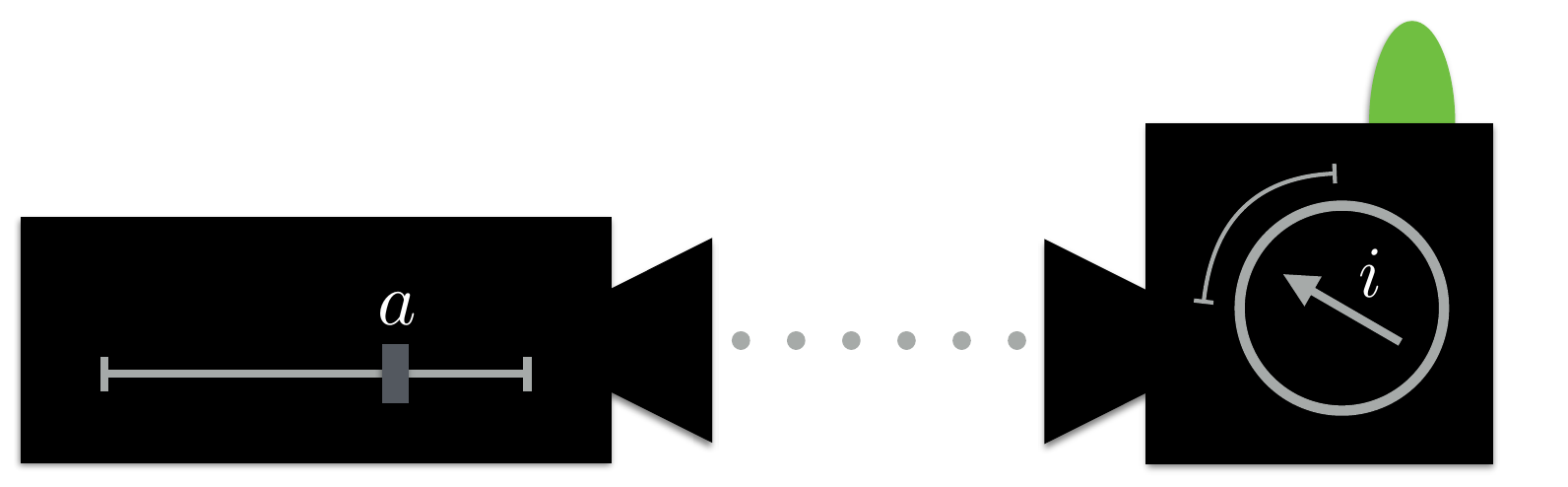}
\caption{
Our state and measurement devices, now with continuous settings!
}\label{cnobs}
\end{figure}

Similarly, as in Figure \ref{demo}, we know what it means to keep the ``state device setting'' constant so that we may coordinate $(p,w)$ \& $(p,v)$ or $(q,w)$ \& $(q,v)$ as being in the same layer.
We even have the notion of an ``average state parameter change'' generated by an ``iso-measurement-ic'' process because we can write
\begin{equation}
	\langle q \rangle = \frac{\langle q w \rangle}{\langle p w \rangle}\langle p \rangle
	\hspace{20pt}
	\text{or}
	\hspace{20pt}
	\langle q \rangle = \frac{\langle q v \rangle}{\langle p v \rangle}\langle p \rangle.
\end{equation}
Further, such an ``average state parameter change'' may not be holonomic because one could have
\begin{equation}
	\frac{\langle q w \rangle}{\langle p w \rangle}
	\neq
	\frac{\langle q v \rangle}{\langle p v \rangle}
\end{equation}
so that the response in the ``average state parameter'' with respect to changes in the ``state device setting'' is a function of ``measurement device setting.''
Importantly, the isomorphism from the ideal piston to SPAM tomography is algebraically exponential | that is, for example,
\begin{equation}
	\frac{\langle q w \rangle}{\langle p w \rangle}
	\sim
	\exp{\!\int\!\dbar Q}.
\end{equation}

\begin{widetext}

\begin{table}[h!]
\vspace{-10pt}
\begin{tabular}{c|c}
Ideal Piston & Toy SPAM Tomography\\\hline
State Space $(S,V)$ & Device Setting Space $(a,i)$\\
Entropy, $S$ & ``State Device Setting'', $a$\\
Volume, $V$ & ``Measurement Device Setting'', $i$\\
Energy, $U(S,V)$ & Data, $\log {\langle pw \rangle_a}^i$\\
Temperature, $T = \left.\frac{\partial U}{\partial S}\right|_V$ & Response, $\chi = \left.\frac{\partial}{\partial a}\right|_i \!\!\log \langle pw \rangle$ \\
Pressure, $P = -\left.\frac{\partial U}{\partial V}\right|_S$ & Response, $\xi = -\left.\frac{\partial}{\partial i}\right|_a \!\log \langle pw \rangle$\\
Heat, $\dbar Q = T dS$ & Average State Parameter Change, $\dbar \log \langle p \rangle = \chi da$\\
Adiabatic/Isentropic & ``Iso-state-ic''\\
Work, $\dbar W = -PdV$ & Average Measurement Parameter Change, $\dbar \log \langle w \rangle = - \xi di$\\
Isochoric & ``Iso-measurement-ic''\\
\end{tabular}
\caption{A table to help with the corresponding terms in the Piston-SPAM analogy.
}\label{correspond}
\end{table}

\end{widetext}

\begin{figure}[h!]
\centering
\includegraphics[width=3in]{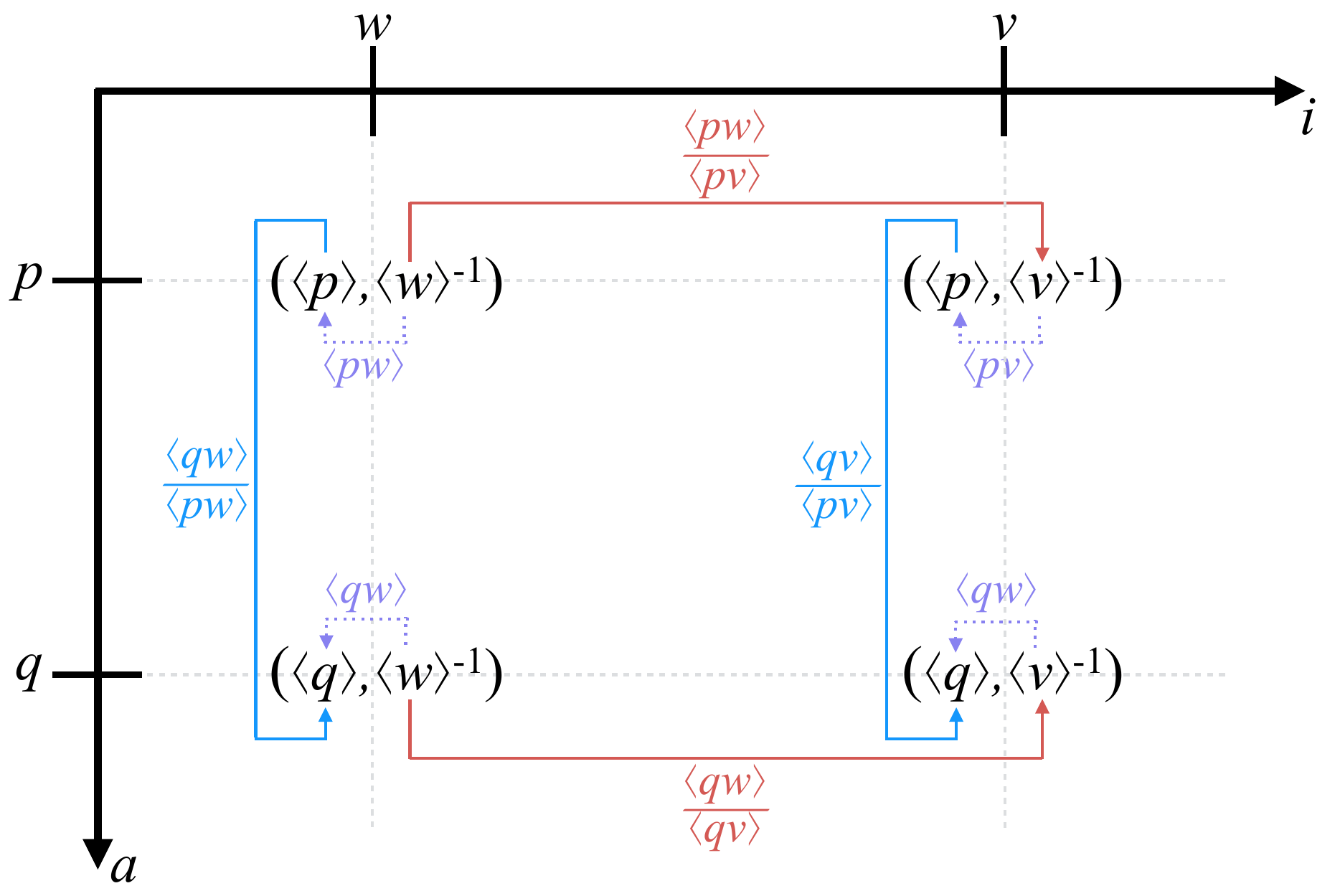}
\caption{An ``S-V'' diagram for toy SPAM tomography. Ratios between horizontally adjacent data can be interpreted as``iso-states-ic'' processes and vertical ratios as ``iso-observables-ic''.  These processes are non-holonomic and so demote the notions of ``average state'' and ``average observable'' from physical coordinates to a gauge degree of freedom.}\label{ThermDat}
\end{figure}

Indeed, this analogy can be made even more exact (see Table \ref{correspond} and Figures \ref{cnobs} and \ref{ThermDat}.)
Returning to our toy devices, suppose instead that the state and observable settings could be dialed continuously and call these external parameters $a$ \& $i$ respectively.
Assuming that $a$ \& $i$ are the only controls, then the data $\langle p w \rangle$ is a well defined function over the space of $(a,i)$.
We can also define responses in the data with respect to these parameters:
\begin{equation}
	\chi = \left.\frac{\partial}{\partial a}\right|_i \!\!\!\log \langle pw \rangle
	\hspace{20pt}
	\text{and}
	\hspace{20pt}
	\xi = -\left.\frac{\partial}{\partial i}\right|_a \!\!\log \langle pw \rangle.
\end{equation}
These responses provide equations of state which we may then attribute to notions of non-holonomic average state parameter \& average measurement parameter changes,
\begin{equation}
	\dbar \log \langle p \rangle = \chi (a,i)d a
	\hspace{20pt}
	\text{and}
	\hspace{20pt}
	\dbar \log \langle w \rangle = -\,\xi (a,i)d i,
\end{equation}
which are related to the original data:
\begin{equation}\label{inexact}
	d \log \langle p w \rangle= \dbar \log \langle p \rangle +\,\dbar \log \langle w \rangle.
\end{equation}
The exponential maps between the finite and the infinitesimal processes may now be written explicitly:
\begin{widetext}

\begin{equation}
	\frac{\langle q w \rangle}{\langle p w \rangle}
	=
	\exp\left(\int_p^q \!\!\chi(a,w) da\right)
	\hspace{20pt}
	\text{and}
	\hspace{20pt}
	\frac{\langle p v  \rangle}{\langle p w \rangle}
	=
	\exp\left(\int_w^v \!\!\xi(p,i) di\right).
\end{equation}

\end{widetext}
Finally, we have for the partial determinant
\begin{widetext}

\begin{equation}
	\Delta = \frac{\langle p v \rangle\langle q w \rangle}{\langle p w \rangle\langle q v \rangle} = \exp\left(\oint \dbar \log \langle p \rangle\right) = \exp\left(-\oint \dbar \log \langle w \rangle\right) = \exp\left(\iint \Gamma\, da di\right)
\end{equation}

\end{widetext}
where the integrals are counterclockwise in Figure \ref{ThermDat} and
\begin{equation}\label{meas?}
	\Gamma = \frac{\partial \chi}{\partial i} =  -\frac{\partial \xi}{\partial a} = -\frac{\partial^2 \log \langle pw \rangle}{\partial a\partial i}
\end{equation}
is a kind of correlation density.

When considering this treatment for the response of quantum data to continuous device settings,
$p$ and $w$ become $d^2 \times d^2$ matrix quantities, $P$ and $W$ such that $D\stackrel{?}=PW$, representing minimally complete tomography experiments for a $d$-dimensional Hilbert space,
as will be explained in section \ref{lattice}.
As such, the inexact forms in Equation \pref{inexact} should be replaced with the forms $(d\langle P \rangle)\langle P \rangle^\inv$ or $\langle W \rangle^\inv d\langle W \rangle$.
These forms may be recognized as Maurer-Cartan forms for the Lie Group $GL(d^2)$ or also the $GL$-equivalent of Mead-Berry Potentials.

\subsection{Analogy: Interactions of a Single Quantum with a Gauge Field}\label{Wilson}

Perhaps the most effective (no pun intended) place to start here is with the gauge interaction of a single electron in an external electromagnetic field.\footnote{Indeed, we could just as well have a discussion about general partition functions in statistical mechanics.
Their dependence on reservoir parameters can be probed with the mode of the ensemble distribution.
The conclusions of such a discussion would have the same essence as the previous section with only the advantage of a technically broader perspective.
The logic would exactly parallel the following discussion so we will not go further than to simply acknowledge its existence.}
The wavefunction can be written as a path integral,
\begin{equation}\label{avgWilson}
	\Psi[\gamma_1,\gamma_0;A] = \int\!\!\mathcal{D}\gamma\, e^{iq\!\int_\gamma \!dx\cdot A}e^{iS_o[\gamma]},
\end{equation}
where $A$ is the 4-vector potential, $S_o$ is the action for the electron in no field,
and the integral is over all paths with initial and final spacetime events $\gamma_0$ \& $\gamma_1$, respectively.
We will not be interested in the spacetime dependence here and will thus denote the wavefunction as just $\Psi[A]$.
On the other hand, the field dependence has a gauge degree of freedom represented by the (projective) symmetry,
\begin{equation}
	\Psi[A+\partial\zeta] = e^{iq(\zeta(\gamma_1)-\zeta(\gamma_0))}\Psi[A]
\end{equation}
which (according to the Born rule) leaves the transition rate between events at $\gamma_0$ \& $\gamma_1$ invariant.
The potential $A$ is also called a \emph{connection} because it fixes the phase of the wavefunction at $\gamma_1$ relative to the phase at $\gamma_0$.

The dominant contribution to the wavefunction is from the path satisfying the classical equation of motion, $\frac{\delta S_o}{\delta \gamma}=qF\dot\gamma$.
If $S_o$ is the free particle action, then the equation of motion is just the Lorentz force law.
However, we could just as well incorporate external interactions into $S_o$ which overpower the Lorentz force and fix $\gamma$ arbitrarily to $\frac{\delta S_o}{\delta \gamma}=0$.
In which case we can write the wavefunction with a classical approximation,
\begin{equation}
	\Psi[\gamma,A] \propto e^{iq\!\int_\gamma \!dx\cdot A}
\end{equation}
where it is understood now that $\gamma$ can be fixed arbitrarily.
We do not bother with the normalization constant or the external phase here because we wish only to illustrate the dependence of the wavefunction on $A$ which we can now imagine is being probed through $\gamma$, which can be externally controlled.

The quantity
\begin{equation}
	W_\gamma = e^{iq\!\int_\gamma \!dx\cdot A}
\end{equation}
is called a Wilson line.
Also important is the Wilson loop
\begin{equation}
	W_\gamma = \Tr \left(e^{iq\!\oint_\gamma \!dx\cdot A}\right)
\end{equation}
where a trace has been introduced to include non-abelian gauge fields where there are several $A$s, one for each generator of the gauge group.
The general wavefunction, Equation \pref{avgWilson}, is often referred to as the ``quantum expectation value'' of the Wilson loop in this context.
Normally, the application of the Wilson loop is to determine the dynamics of $\gamma$ from a theory of the gauge field.
However, our purpose for the Wilson loop is to represent how the gauge field could be probed by an externally fixed $\gamma$.  (See Figure \ref{triad})

When we consider partial determinants in section \ref{lattice}, the analogous quantity will be just the closed Wilson line, i.e. a Wilson loop without the trace.
Aside from the difference between a single number and a matrix,
an important distinction is that closed Wilson line actually depend on the initial/final point from which $\gamma$ is drawn, while Wilson loops do not.
However, the dependence is simple and only such that the closed Wilson line is gauge covariant instead of invariant
\begin{equation}\label{covar}
	W_\gamma \longrightarrow U(\gamma_1)W_\gamma U^\inv(\gamma_0)
\end{equation}
where $\gamma_1=\gamma_0$ for a closed contour.
Although this does not have any significance in gauge field theories, it is significant for a theory of SPAM correlations.

Analogous to a Wilson line, one can define a tomography line:
\begin{equation}\label{taugauge}
\Delta\big(\gamma,\tau\big) = \exp{\!\int_\gamma\!\tau}.
\end{equation}
which represents a specific type tomography, where the gauge parameter of experiment $\gamma_1$ is concluded from the gauge parameter of experiment $\gamma_0$
through the data, represented by the connection $\tau$, along changes in the device parameters, represented by the contour $\gamma$.
The tomographic connection, $\tau$, is not uniquely determined by the data but is nonetheless intimately related to the interpretation of the gauge at each experiment along $\gamma$.
Formally this is represented by the tomography lines being equivalent by a local gauge transformation
\begin{equation}
\Delta\big(\gamma,\tau+dg\big) = e^{g(\gamma_1)-g(\gamma_0)}\Delta\big(\gamma,\tau)= e^{g(\gamma_1)}\Delta\big(\gamma,\tau)e^{-g(\gamma_0)}
\end{equation}
where the effect of the transformation is only to relabel the initial and final gauge parameters.

Returning to our toy devices, suppose that the gauge at each experiment is represented by an average state parameter (one could call this fixing the \emph{state} gauge.)
Then for $da=0$, $\Delta$ would be the identity, while along the $a$-direction
\begin{equation}\label{sgauge}
\Delta\big(\gamma,\dbar \log\langle p \rangle\big) = \exp\left(\int_\gamma\!\chi da\right)
\end{equation}
would represent iso-measurement-ic tomography.
Similarly, if the gauge at each experiment is represented by an average measurement parameter (let's call this \emph{measurement} gauge), then
\begin{equation}\label{mgauge}
\Delta\big(\gamma,-\,\dbar \log\langle w \rangle\big) = \exp\left(\int_\gamma\! \xi di\right)
\end{equation}
would represent iso-state-ic tomography.
Most importantly, these tomographies are equivalent to each other modulo a local gauge transformation:
\begin{align}
\Delta\big(\gamma,-\,\dbar \log\langle w \rangle\big) & = \Delta\big(\gamma,\,\dbar \log\langle p \rangle - d\log\langle pw\rangle\big)\\
& = \exp\left(-\!\int_\gamma\!d \log\langle pw \rangle\right)\Delta\big(\gamma,\,\dbar \log\langle p \rangle\big)\\
& = \frac{\langle pw \rangle(\gamma_0)}{\langle pw \rangle(\gamma_1)}\Delta\big(\gamma,\,\dbar \log\langle p \rangle\big).
\end{align}
In the electromagnetism analogy, these are the equivalent of Landau gauges (see Figure \ref{landau}.)

\subsection{Non-Holonomic Quantum Tomography and Non-Abelian Lattice Gauge}\label{lattice}

Having hopefully made the perspective of non-holonomic tomography clear through these analogies for the toy problem,
some discussion about the actual quantum problem is due.\cite{jackson2015detecting}
The quantum problem is the same as the toy problem
except that we assume the state and measurement devices are parameterized by Hermitian operators (a density operator and a POVM element, respectively) over a $d$-dimensional Hilbert space.
In particular, this means that the devices are to be modeled by $d^2$ random variables each.
If all the device parameters were uncorrelated, then one could write these operators as
\begin{equation}
\rho_a = \frac{1}{d}{p_a}^\mu\sigma_\mu
\hspace{25pt}\text{and}\hspace{25pt}
E^i = \sigma^\mu{w_\mu}^i
\end{equation}
where the $\{\sigma_\mu\}_{\mu=0}^{d^2-1}$ is some operator basis of Hermitian operators, $\{\sigma^\mu\}$ is its reciprocal basis, and a sum over repeated indices is implied.
If $\sigma_0=1$ and the other $\sigma_\mu$ are traceless, then ${p_a}^0$ and ${w_0}^i$ are identical to the single device parameters of the toy problem.

The measured frequencies, a.k.a. ``the data'',  are now given by
\begin{equation}
	{f_a}^i = {\langle \Tr \rho E\rangle_a}^i = {\langle p^\mu w_\mu \rangle_a}^i.
\end{equation}
To be effectively uncorrelated in this case means that the data can be decomposed into the form
\begin{widetext}

\begin{equation}\label{quanteff}
F=
\left[
\begin{array}{ccc}
	{\langle p^\mu w_\mu \rangle_1}^1	&&	{\langle p^\mu w_\mu \rangle_1}^M	\\
		&	\ddots	&	\\
	{\langle p^\mu w_\mu \rangle_N}^1	&&	{\langle p^\mu w_\mu \rangle_N}^M	
\end{array}
\right]
=
\left[
\begin{array}{ccc}
	\langle p^0 \rangle_1 & \cdots & \langle p^{d^2-1} \rangle_1	\\
	&\vdots&	\\
	\langle p^0 \rangle_N & \cdots & \langle p^{d^2-1} \rangle_N	
\end{array}
\right]
\left[
\begin{array}{ccc}
	\langle w_0 \rangle^1	& &	\langle w_{d^2-1} \rangle^M\\
	\vdots&\cdots&\vdots\\
	\langle w_0 \rangle^1	& &	\langle w_{d^2-1} \rangle^M
\end{array}
\right],
\end{equation}

\end{widetext}
which is equivalent to saying that the rank is bounded above by $\rank(D) \le d^2$.
To define a partial determinant, the simplest way is to consider $M=N=2d^2$ and partition the data into 4 $d^2\!\times\!d^2$ corners,
\begin{equation}\label{corners}
F=
\left[
\begin{array}{cc}
	A & B \\
	C & D
\end{array}
\right].
\end{equation}
The partial determinant is
\begin{equation}\label{QPD}
	\Delta(F) = D^\inv C B^\inv A
\end{equation}
which is significant because of the result
\begin{verse}\centering
	F is globally uncorrelated if and only if $\Delta(F)=1$.
\end{verse}
Specifically, $\Delta$ parameterizes $d^4$ degrees of correlation.
However, because of gauge covariance (Equation \ref{covar},) only $d^2$ of these are gauge invariant parameters.

\begin{figure}[h!]
\centering
\includegraphics[height=0.75in]{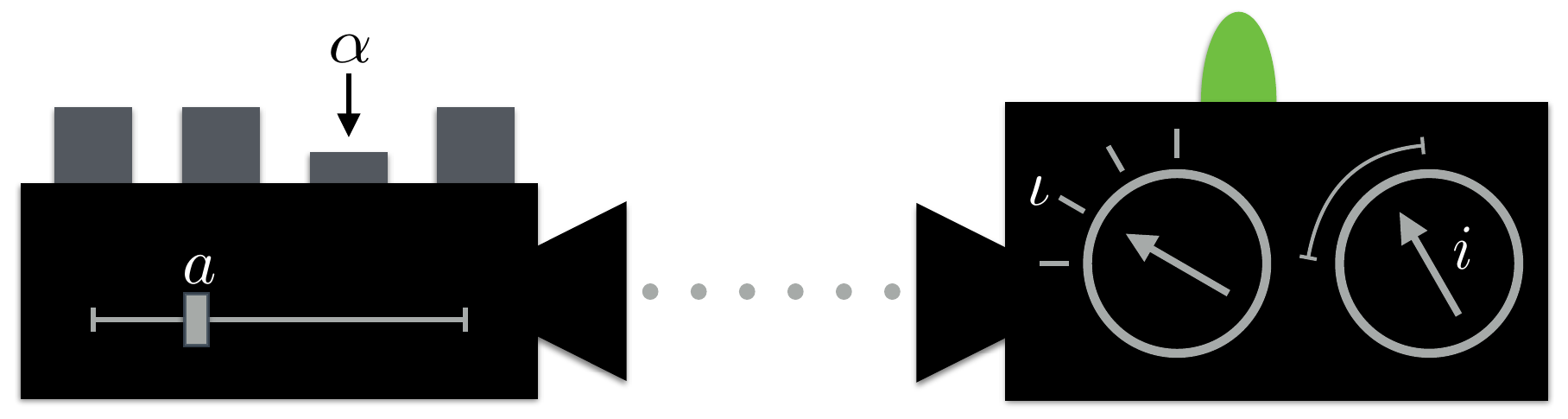}
\caption{
The $d^2$ buttons enumerate a (detector) tomographically complete frame of states.
The $d^2$ notches enumerate a (state) tomographically complete frame of observables.
The continuous slider and continuous dial are the square coordinates which displace settings.
}\label{cknobs}
\end{figure}

In the quantum case, it becomes important to pay attention to the arrangement of the settings when the data is considered in the form of Equation \pref{corners} so let us define indices:
\begin{equation}
F
=
\left[
\begin{array}{cc}
	A & B \\
	C & D
\end{array}
\right]
=
\left[
\begin{array}{cc}
	{D_0}^0 & {D_0}^1 \\
	{D_1}^0 & {D_1}^1
\end{array}
\right]
\end{equation}
where the matrix elements of these corners are
\begin{equation}
{({D_a}^i)_\alpha}^\iota
=
{f_{ad^2+\alpha+1}}^{id^2+\iota+1}.
\end{equation}
The \emph{corners} are coordinated by $(a,i)$ and understood to be $2\times2$ minimally complete tomography experiments we call a \emph{square}.
Each minimally complete tomography experiment consists of $d^2$ states enumerated by $\alpha$ and $d^2$ measurements enumerated by $\iota$
and is further associated with $d^4$ gauge degrees of freedom reflecting the fact that the data of each corner is locally (effectively) uncorrelated,
\begin{equation}\label{stuff}
	{({D_a}^i)_\alpha}^\iota = {(P_a)_\alpha}^\mu{(W^i)_\mu}^\iota
	= {(P_a)_\alpha}^\mu {G_\mu}^\lambda {G^\inv_{\,\lambda}}^\nu {(W^i)_\nu}^\iota.
\end{equation}
The corners are understood to be displaced from each other through changes in $(a,i)$
and it is useful to think of these indices as pairs of points on a continuum (see Figure \ref{cknobs}.)
As such, the data matrix is conceptually reorganized as a square which has gauges at each corner (experiment)
which are connected to each other by the edges over which the data define a connection.

For simplicity, each minimally complete tomography experiment will henceforth be referred to as just an experiment.
For each experiment, the Born rule, $A=PW$, can be thought of as a connection between gauge parameters, e.g. $P = A W^\inv$ or $W^\inv\xrightarrow{A}P$.
In other words, the data from experiments can be interpreted as defining maps.
For multiple experiments sharing devices, there are degrees of choice as to how one can represent the gauge degrees of freedom for each pair of devices.
These choices simultaneously correspond to the choices of how to embed the data in the maps between these experiments.
Let us go over a few particularly meaningful examples.

A couple of gauges that should be familiar are what we would like to call standard gauges (Figure \ref{standard}.)
Every arrow represents a constraint which may be interpreted as a tomography | e.g. in the right diagram of Figure \ref{standard}, $P\xrightarrow{A^\inv}W^\inv$ represents the equation $W^\inv = A^\inv P$ which may be interpreted as a detector tomography.
This gauge is in fact the gauge used in the tomographic proof of section \ref{PD}.
Also important are what we call tomographies in ``Landau'' gauge (see Figure \ref{landau}) which have actually appeared (sections \ref{PD} and \ref{OMGthermo}.)
The reader is encouraged to stare at these 4 gauges and try to see how they are each an equivalent representation of the same organization of information as Figure \ref{ThermDat}.
 
\begin{figure}[H]\centering
\begin{tikzpicture}[>=angle 90]
\matrix(a)[matrix of math nodes,row sep=3em, column sep=3em,text height=1.5ex, text depth=0.25ex]{W^\inv&P &&& P& V^\inv\\ Q & V^\inv &&& W^\inv & Q \\};
\path[->] (a-1-1) edge node[above]{$A$} (a-1-2);
\path[->] (a-1-1) edge node[left]{$C$} (a-2-1);
\path[->] (a-1-2) edge node[right]{$B^\inv$} (a-2-2);
\path[->] (a-2-1) edge node[below]{$D^\inv$} (a-2-2);
\path[->] (a-1-5) edge node[left]{$A^\inv$} (a-2-5);
\path[->] (a-1-5) edge node[above]{$B^\inv$} (a-1-6);
\path[->] (a-1-6) edge node[right]{$D$} (a-2-6);
\path[->] (a-2-5) edge node[below]{$C$} (a-2-6);
\end{tikzpicture}
\caption{Tomography in ``Standard'' Gauge.  We call them standard gauges because, considering for instance the left connection:
The measurement parameters of the top-left experiment are imagined to be fixed
in which case the data from this experiment can be interpreted as a standard state tomography on the top-right experiment, and from the top-right the connection does standard detector tomography on the bottom-right, etc.
The choice of representing the top-left experiment's gauge by its measurement device parameters, the top-right experiment's gauge by its state device paramters, etc.
uniquely defines how the data is to be organized as a connection in between these experiment's gauge parameters.
\label{standard}}
\end{figure}

\begin{figure}[H]\centering
\begin{tikzpicture}[>=angle 90]
\matrix(a)[matrix of math nodes,row sep=3em, column sep=3em,text height=1.5ex, text depth=0.25ex]{P& P &&&  W^\inv&V^\inv \\ Q & Q &&& W^\inv & V^\inv  \\};
\path[->] (a-1-1) edge node[above]{$1$} (a-1-2);
\path[->] (a-1-1) edge node[left]{$C A^\inv$} (a-2-1);
\path[->] (a-1-2) edge node[right]{$D B^\inv$} (a-2-2);
\path[->] (a-2-1) edge node[below]{$1$} (a-2-2);
\path[->] (a-1-5) edge node[above]{$B^\inv A$} (a-1-6);
\path[->] (a-1-5) edge node[left]{$1$} (a-2-5);
\path[->] (a-1-6) edge node[right]{$1$} (a-2-6);
\path[->] (a-2-5) edge node[below]{$D^\inv C$} (a-2-6);
\end{tikzpicture}
\caption{Tomography in ``Landau'' Gauge.
Left: iso-measurement-ic tomography, the arrangement of quantum data in state gauge, Equation \pref{sgauge}.
Right: iso-state-ic tomography, the arrangement of quantum data in measurement gauge, Equation \pref{mgauge}.
These are called Landau because they keep gauge parameters in either the state or measurement direction constant
just like the vector potential for a 2-d surface in the x- or y-direction can be chosen to be zero .
The left gauge is a tomography where data from two experiments (either A and C or B and D) with a common measurement device
is used to infer an unknown state device (Q) from a ``known'' state device (P.) 
This kind of tomography has been thought of before and already put into practice \cite{cooper2014local} (instead using a maximum likelihood method to estimate parameters rather than linear inversion, which we are considering.)
As far as the authors are aware, the right gauge is a tomography yet unperformed.
\label{landau}}
\end{figure}

\begin{figure}[H]\centering
\begin{tikzpicture}[>=angle 90]
\matrix(a)[matrix of math nodes,row sep=3em, column sep=3em,text height=1.5ex, text depth=0.25ex]{
\square & \square & \square && \square & \square & \square \\
\square & \Gamma & \square && \square &  g\Gamma & \square \\
\square &\square & \square && \square & \square & \square \\};
\path[->,dotted] (a-1-1) edge (a-1-2);	\path[->,dotted] (a-1-2) edge (a-1-3);
\path[->,dotted] (a-1-1) edge (a-2-1);	\path[->,dotted] (a-1-3) edge (a-2-3);
\path[->,dotted] (a-2-1) edge (a-3-1);	\path[->,dotted] (a-2-3) edge (a-3-3);
\path[->,dotted] (a-3-1) edge (a-3-2);	\path[->,dotted] (a-3-2) edge (a-3-3);
\path[->] (a-2-2) edge node[below]{$X$} (a-2-3);
\path[->] (a-2-2) edge node[left]{$Y$} (a-3-2);
\path[->] (a-1-2) edge node[right]{$T$} (a-2-2);
\path[->] (a-2-1) edge node[above]{$Z$} (a-2-2);
\path[->,dotted] (a-1-5) edge (a-1-6);	\path[->,dotted] (a-1-6) edge (a-1-7);
\path[->,dotted] (a-1-5) edge (a-2-5);	\path[->,dotted] (a-1-7) edge (a-2-7);
\path[->,dotted] (a-2-5) edge (a-3-5);	\path[->,dotted] (a-2-7) edge (a-3-7);
\path[->,dotted] (a-3-5) edge (a-3-6);	\path[->,dotted] (a-3-6) edge (a-3-7);
\path[->] (a-2-6) edge node[left]{$Y g^\inv$} (a-3-6);
\path[->] (a-2-6) edge node[below]{$X g^\inv$} (a-2-7);
\path[->] (a-1-6) edge node[right]{$g T$} (a-2-6);
\path[->] (a-2-5) edge node[above]{$g Z$} (a-2-6);
\end{tikzpicture}
\caption{Local Gauge Transformations:
The vertical direction represents displacements in state $a$ and the horizontal direction represents displacements in measurement $i$.
At each vertex (experiment) is a $d^2 \times d^2$ matrix of gauge parameters, $\Gamma$.
At each adjacent edge (connection to the adjacent experiment) is a component of the connection,
$X=\tau_i(a,i)$, $Y=\tau_a(a,i)$, $Z=\tau_a(a-1,i)$, $T=\tau_i(a,i-1)$ (see Equation \ref{taugauge}.)
The distance between lattice sites is defined by distances along continuous device settings (see Figure \ref{cknobs}.)
The right lattice is a gauge transformation, g, of the left lattice at just the one vertex.
These transformations leave the constraints represented by each connection invariant.
\label{lattgaug}}
\end{figure}
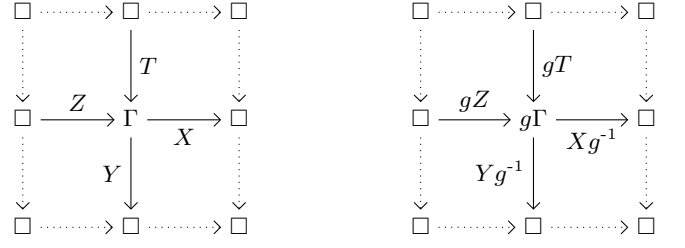

All of these gauges are formally related to each other by local gauge transformations.
As such, an explanation of gauge transformations on a lattice is in order (see Figure \ref{lattgaug}.)
Instead of considering only a square of experiments,
it is conceptually more useful to think about a \emph{lattice} of experiments sharing devices.
Something to notice is that $g$ is not exactly the $G$ in Equation \ref{stuff}, but rather $g\Gamma = \Gamma G$ or $g = \Gamma G \Gamma^\inv$.

Having re-expressed non-holonomic tomography for quantum systems, some distinctions are in order.
First, as already mentioned one should not forget that unlike in the toy model, the gauges of quantum tomography are non-abelian
| particularly, the gauge does not generally commute with the connection
| which results in a covariance (see Equation \ref{covar}) of closed-line tomographies on the gauge at the initial/terminal experiment.
Second, the gauge groups, $\mathrm{GL}(d^2,\R)$, we are concerned with are actually not compact like the unitary groups of Yang-Mills theories.\footnote{
.. ignoring positivity constraints.}
Third, one could imagine having $d^2$ continuous settings per device,
in which case the gauge group becomes a tangent space, where the frame, $P$, and coframe, $W^\inv$, are then like vierbein.
Fourth, an experimentalist may not have any ``sliders'' but rather just have $2d^2$ ``buttons'' per device
in which case a metric for the distance between experiments is obscured.
Finally, in the ``only buttons'' scenario, localizing settings to corners of a square becomes arbitrary
| i.e. whether settings $\{1,2,3,4\}$ are to appear in the first corner or $\{2,6,4,7\}$ is arbitrary.

\section{Conclusion and Discussion}

In this work, we considered non-holonomic quantum tomography as a perspective for the method of partial determinants\cite{jackson2015detecting}.
Partial determinants are matrix quantities which analyze quantum data to detect and quantify SPAM correlations,
without estimating average state-preparation or measurement parameters.
We particularly focused on a toy model to illustrate that the partial determinant is in fact a holonomy,
showing that one can formalize SPAM tomographies in direct analogy to thermodynamic theories and gauge field theories.
A SPAM tomography is then non-holonomic if the partial determinants (i.e. tomographic holonomies) have nontrivial values,
which can be interpreted as correlations between state and measurement parameters.

\begin{figure}[h!]
\centering
\includegraphics[height=1.25in]{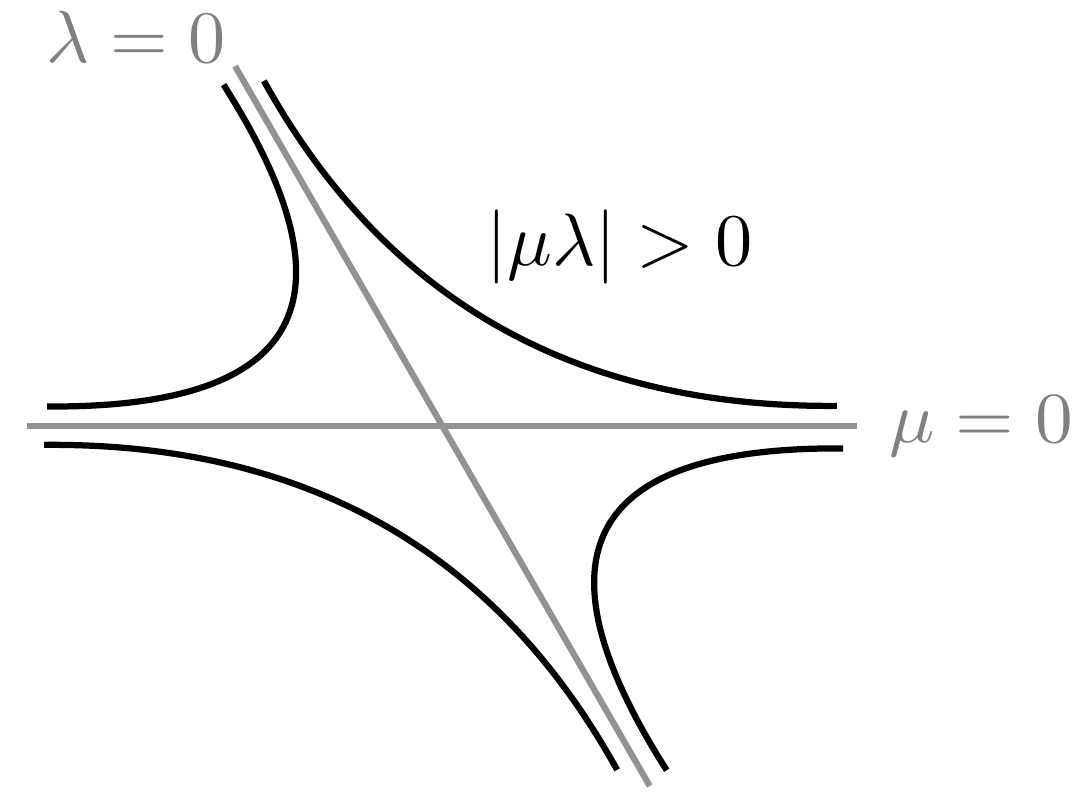}
\caption{
Using a determinant to define the distance of a rank 2 matrix from the space of rank 1 matrices can be a subtle point.
If $\lambda$ and $\mu$ are the singular values of a matrix $M$, then $|\Det M| = \lambda\mu$ is a type of distance from the axes (which are rank 1), modulo area preserving transformations.
The axes are drawn askew to emphasize that there is no notion of metric distance.
}\label{product}
\end{figure}

From a practical perspective, the matrix elements of a PD can be used to \emph{detect} amounts of SPAM correlation.
However, the way in which a PD measures distances away from a correlated model can be a little subtle because these distances are not a metric, in the standard mathematical sense.
The subtlety simply reduces to the fact that the determinant of a matrix alone does not actually tell you how large its smallest singular value is (see Figure \ref{product}.)
Rather than think of distances away from the space of uncorrelated data, one must think in terms of inherited notions of distance from continuous device settings.
The equations such as \pref{meas?} can quantitatively measure correlations relative to areas in setting space.

A broader observation should also be made about device parameters and gauge dimensions.
Importantly, one should notice that the only property which distinguished the toy problem from the quantum problem
was a mere ``speculation'' about the number of degrees of freedom which parameterize the devices.
In the most general scheme, an $r \times r$ PD is a test of the ability to model the data by uncorrelated $r$-dimensional state and measurement vectors.
For quantum probabilities, one has further interpretations for the $r=d^2$ dimensions
reflecting that the state and measurement vectors are also operators on a $d$-dimensional vector space.
As an example of a more general application of PDs, one could consider $2\times2$ PDs for an uncorrelated qubit system.
Such a PD would generally take a value different from the identity which can be interpreted as a measure of the inability to model the data by uncorrelated \emph{classical bit} state and measurement parameters.

\begin{acknowledgments}
This research was supported in part by Perimeter Institute for Theoretical Physics.
Research at Perimeter Institute is supported by the Government of Canada through the Department of Innovation, Science and Economic Development
and by the Province of Ontario through the Ministry of Research, Innovation and Science.
C. Jackson would like to particularly thank R. Spekkens for a useful conversation which helped to make section \ref{toy} more concrete.
S.J. van Enk was supported in part by ARO/LPS under Contract No. W911NF-14-C-0048.
\end{acknowledgments}

\bibliography{Non-Holo_Persp}

\begin{thebibliography}{16}%
\makeatletter
\providecommand \@ifxundefined [1]{%
 \@ifx{#1\undefined}
}%
\providecommand \@ifnum [1]{%
 \ifnum #1\expandafter \@firstoftwo
 \else \expandafter \@secondoftwo
 \fi
}%
\providecommand \@ifx [1]{%
 \ifx #1\expandafter \@firstoftwo
 \else \expandafter \@secondoftwo
 \fi
}%
\providecommand \natexlab [1]{#1}%
\providecommand \enquote  [1]{``#1''}%
\providecommand \bibnamefont  [1]{#1}%
\providecommand \bibfnamefont [1]{#1}%
\providecommand \citenamefont [1]{#1}%
\providecommand \href@noop [0]{\@secondoftwo}%
\providecommand \href [0]{\begingroup \@sanitize@url \@href}%
\providecommand \@href[1]{\@@startlink{#1}\@@href}%
\providecommand \@@href[1]{\endgroup#1\@@endlink}%
\providecommand \@sanitize@url [0]{\catcode `\\12\catcode `\$12\catcode
  `\&12\catcode `\#12\catcode `\^12\catcode `\_12\catcode `\%12\relax}%
\providecommand \@@startlink[1]{}%
\providecommand \@@endlink[0]{}%
\providecommand \url  [0]{\begingroup\@sanitize@url \@url }%
\providecommand \@url [1]{\endgroup\@href {#1}{\urlprefix }}%
\providecommand \urlprefix  [0]{URL }%
\providecommand \Eprint [0]{\href }%
\providecommand \doibase [0]{http://dx.doi.org/}%
\providecommand \selectlanguage [0]{\@gobble}%
\providecommand \bibinfo  [0]{\@secondoftwo}%
\providecommand \bibfield  [0]{\@secondoftwo}%
\providecommand \translation [1]{[#1]}%
\providecommand \BibitemOpen [0]{}%
\providecommand \bibitemStop [0]{}%
\providecommand \bibitemNoStop [0]{.\EOS\space}%
\providecommand \EOS [0]{\spacefactor3000\relax}%
\providecommand \BibitemShut  [1]{\csname bibitem#1\endcsname}%
\let\auto@bib@innerbib\@empty
\bibitem [{\citenamefont {Jackson}\ and\ \citenamefont {van
  Enk}(2015)}]{jackson2015detecting}%
  \BibitemOpen
  \bibfield  {author} {\bibinfo {author} {\bibfnamefont {C.}~\bibnamefont
  {Jackson}}\ and\ \bibinfo {author} {\bibfnamefont {S.~J.}\ \bibnamefont {van
  Enk}},\ }\href@noop {} {\bibfield  {journal} {\bibinfo  {journal} {Physical
  Review A}\ }\textbf {\bibinfo {volume} {92}},\ \bibinfo {pages} {042312}
  (\bibinfo {year} {2015})}\BibitemShut {NoStop}%
\bibitem [{\citenamefont {Jackson}\ and\ \citenamefont {van
  Enk}(2017)}]{nonholo2}%
  \BibitemOpen
  \bibfield  {author} {\bibinfo {author} {\bibfnamefont {C.}~\bibnamefont
  {Jackson}}\ and\ \bibinfo {author} {\bibfnamefont {S.}~\bibnamefont {van
  Enk}},\ }\href@noop {} {\bibfield  {journal} {\bibinfo  {journal} {arXiv
  preprint arXiv:1702.06090}\ } (\bibinfo {year} {2017})}\BibitemShut {NoStop}%
\bibitem [{\citenamefont {Merkel}\ \emph {et~al.}(2013)\citenamefont {Merkel},
  \citenamefont {Gambetta}, \citenamefont {Smolin}, \citenamefont {Poletto},
  \citenamefont {C{\'o}rcoles}, \citenamefont {Johnson}, \citenamefont {Ryan},\
  and\ \citenamefont {Steffen}}]{merkel}%
  \BibitemOpen
  \bibfield  {author} {\bibinfo {author} {\bibfnamefont {S.~T.}\ \bibnamefont
  {Merkel}}, \bibinfo {author} {\bibfnamefont {J.~M.}\ \bibnamefont
  {Gambetta}}, \bibinfo {author} {\bibfnamefont {J.~A.}\ \bibnamefont
  {Smolin}}, \bibinfo {author} {\bibfnamefont {S.}~\bibnamefont {Poletto}},
  \bibinfo {author} {\bibfnamefont {A.~D.}\ \bibnamefont {C{\'o}rcoles}},
  \bibinfo {author} {\bibfnamefont {B.~R.}\ \bibnamefont {Johnson}}, \bibinfo
  {author} {\bibfnamefont {C.~A.}\ \bibnamefont {Ryan}}, \ and\ \bibinfo
  {author} {\bibfnamefont {M.}~\bibnamefont {Steffen}},\ }\href@noop {}
  {\bibfield  {journal} {\bibinfo  {journal} {Phys. Rev. A}\ }\textbf {\bibinfo
  {volume} {87}},\ \bibinfo {pages} {062119} (\bibinfo {year}
  {2013})}\BibitemShut {NoStop}%
\bibitem [{\citenamefont {Blume-Kohout}\ \emph {et~al.}(2013)\citenamefont
  {Blume-Kohout}, \citenamefont {Gamble}, \citenamefont {Nielsen},
  \citenamefont {Mizrahi}, \citenamefont {Sterk},\ and\ \citenamefont
  {Maunz}}]{gst}%
  \BibitemOpen
  \bibfield  {author} {\bibinfo {author} {\bibfnamefont {R.}~\bibnamefont
  {Blume-Kohout}}, \bibinfo {author} {\bibfnamefont {J.~K.}\ \bibnamefont
  {Gamble}}, \bibinfo {author} {\bibfnamefont {E.}~\bibnamefont {Nielsen}},
  \bibinfo {author} {\bibfnamefont {J.}~\bibnamefont {Mizrahi}}, \bibinfo
  {author} {\bibfnamefont {J.~D.}\ \bibnamefont {Sterk}}, \ and\ \bibinfo
  {author} {\bibfnamefont {P.}~\bibnamefont {Maunz}},\ }\href@noop {}
  {\bibfield  {journal} {\bibinfo  {journal} {arXiv preprint arXiv:1310.4492}\
  } (\bibinfo {year} {2013})}\BibitemShut {NoStop}%
\bibitem [{\citenamefont {Stark}(2014)}]{stark}%
  \BibitemOpen
  \bibfield  {author} {\bibinfo {author} {\bibfnamefont {C.}~\bibnamefont
  {Stark}},\ }\href@noop {} {\bibfield  {journal} {\bibinfo  {journal} {Phys.
  Rev. A}\ }\textbf {\bibinfo {volume} {89}},\ \bibinfo {pages} {052109}
  (\bibinfo {year} {2014})}\BibitemShut {NoStop}%
\bibitem [{Note1()}]{Note1}%
  \BibitemOpen
  \bibinfo {note} {There is a slight conflict of language here as modern field
  theorists like to use ``observable'' to refer to cross-sections, lifetimes,
  etc. which we refer to as ``statistical observables'' as opposed to ``quantum
  observables'' which refer to operators in a theory and what we mean
  throughout this paper by ``observable.''}\BibitemShut {NoStop}%
\bibitem [{\citenamefont {Lvovsky}\ and\ \citenamefont
  {Raymer}(2009)}]{raymer}%
  \BibitemOpen
  \bibfield  {author} {\bibinfo {author} {\bibfnamefont {A.~I.}\ \bibnamefont
  {Lvovsky}}\ and\ \bibinfo {author} {\bibfnamefont {M.~G.}\ \bibnamefont
  {Raymer}},\ }\href@noop {} {\bibfield  {journal} {\bibinfo  {journal} {Rev.
  Mod. Phys.}\ }\textbf {\bibinfo {volume} {81}},\ \bibinfo {pages} {299}
  (\bibinfo {year} {2009})}\BibitemShut {NoStop}%
\bibitem [{\citenamefont {Paris}\ and\ \citenamefont {Rehacek}(2004)}]{paris}%
  \BibitemOpen
  \bibfield  {author} {\bibinfo {author} {\bibfnamefont {M.}~\bibnamefont
  {Paris}}\ and\ \bibinfo {author} {\bibfnamefont {J.}~\bibnamefont
  {Rehacek}},\ }\href@noop {} {\emph {\bibinfo {title} {Quantum state
  estimation}}},\ Vol.\ \bibinfo {volume} {649}\ (\bibinfo  {publisher}
  {Springer Science \& Business Media},\ \bibinfo {year} {2004})\BibitemShut
  {NoStop}%
\bibitem [{\citenamefont {Lundeen}\ \emph {et~al.}(2009)\citenamefont
  {Lundeen}, \citenamefont {Feito}, \citenamefont {Coldenstrodt-Ronge},
  \citenamefont {Pregnell}, \citenamefont {Silberhorn}, \citenamefont {Ralph},
  \citenamefont {Eisert}, \citenamefont {Plenio},\ and\ \citenamefont
  {Walmsley}}]{lundeen}%
  \BibitemOpen
  \bibfield  {author} {\bibinfo {author} {\bibfnamefont {J.}~\bibnamefont
  {Lundeen}}, \bibinfo {author} {\bibfnamefont {A.}~\bibnamefont {Feito}},
  \bibinfo {author} {\bibfnamefont {H.}~\bibnamefont {Coldenstrodt-Ronge}},
  \bibinfo {author} {\bibfnamefont {K.}~\bibnamefont {Pregnell}}, \bibinfo
  {author} {\bibfnamefont {C.}~\bibnamefont {Silberhorn}}, \bibinfo {author}
  {\bibfnamefont {T.}~\bibnamefont {Ralph}}, \bibinfo {author} {\bibfnamefont
  {J.}~\bibnamefont {Eisert}}, \bibinfo {author} {\bibfnamefont
  {M.}~\bibnamefont {Plenio}}, \ and\ \bibinfo {author} {\bibfnamefont
  {I.}~\bibnamefont {Walmsley}},\ }\href@noop {} {\bibfield  {journal}
  {\bibinfo  {journal} {Nature Physics}\ }\textbf {\bibinfo {volume} {5}},\
  \bibinfo {pages} {27} (\bibinfo {year} {2009})}\BibitemShut {NoStop}%
\bibitem [{\citenamefont {Feito}\ \emph {et~al.}(2009)\citenamefont {Feito},
  \citenamefont {Lundeen}, \citenamefont {Coldenstrodt-Ronge}, \citenamefont
  {Eisert}, \citenamefont {Plenio},\ and\ \citenamefont {Walmsley}}]{feito}%
  \BibitemOpen
  \bibfield  {author} {\bibinfo {author} {\bibfnamefont {A.}~\bibnamefont
  {Feito}}, \bibinfo {author} {\bibfnamefont {J.}~\bibnamefont {Lundeen}},
  \bibinfo {author} {\bibfnamefont {H.}~\bibnamefont {Coldenstrodt-Ronge}},
  \bibinfo {author} {\bibfnamefont {J.}~\bibnamefont {Eisert}}, \bibinfo
  {author} {\bibfnamefont {M.}~\bibnamefont {Plenio}}, \ and\ \bibinfo {author}
  {\bibfnamefont {I.}~\bibnamefont {Walmsley}},\ }\href@noop {} {\bibfield
  {journal} {\bibinfo  {journal} {New Journal of Physics}\ }\textbf {\bibinfo
  {volume} {11}},\ \bibinfo {pages} {093038} (\bibinfo {year}
  {2009})}\BibitemShut {NoStop}%
\bibitem [{\citenamefont {Cooper}\ \emph {et~al.}(2014)\citenamefont {Cooper},
  \citenamefont {Karpi{\'n}ski},\ and\ \citenamefont
  {Smith}}]{cooper2014local}%
  \BibitemOpen
  \bibfield  {author} {\bibinfo {author} {\bibfnamefont {M.}~\bibnamefont
  {Cooper}}, \bibinfo {author} {\bibfnamefont {M.}~\bibnamefont
  {Karpi{\'n}ski}}, \ and\ \bibinfo {author} {\bibfnamefont {B.~J.}\
  \bibnamefont {Smith}},\ }\href@noop {} {\bibfield  {journal} {\bibinfo
  {journal} {Nature communications}\ }\textbf {\bibinfo {volume} {5}} (\bibinfo
  {year} {2014})}\BibitemShut {NoStop}%
\bibitem [{Note2()}]{Note2}%
  \BibitemOpen
  \bibinfo {note} {Of course, $g$ must has a compact range so that the
  interpretation of $\big (g\langle p\rangle , g^\inv \langle w\rangle \big )$
  as a pair of probabilities still makes sense. However, this detail is not of
  concern for this paper.}\BibitemShut {Stop}%
\bibitem [{Note3()}]{Note3}%
  \BibitemOpen
  \bibinfo {note} {One can argue that such devices do not exist other than by
  assumption!}\BibitemShut {Stop}%
\bibitem [{Note4()}]{Note4}%
  \BibitemOpen
  \bibinfo {note} {Remember that this result is equivalent to the commonly
  known property that $D$ is rank ($\le $) 1 if and only if $\protect \mathrm
  {Det}D = 0$.}\BibitemShut {Stop}%
\bibitem [{Note5()}]{Note5}%
  \BibitemOpen
  \bibinfo {note} {Indeed, we could just as well have a discussion about
  general partition functions in statistical mechanics. Their dependence on
  reservoir parameters can be probed with the mode of the ensemble
  distribution. The conclusions of such a discussion would have the same
  essence as the previous section with only the advantage of a technically
  broader perspective. The logic would exactly parallel the following
  discussion so we will not go further than to simply acknowledge its
  existence.}\BibitemShut {Stop}%
\bibitem [{Note6()}]{Note6}%
  \BibitemOpen
  \bibinfo {note} {.. ignoring positivity constraints.}\BibitemShut {Stop}%
\end{thebibliography}%
\end{document}